\documentclass[%
 aps,
 superscriptaddress,
 nofootinbib,
 amsmath,amssymb,
 aps,
 prb,
 twocolumn,
]{revtex4-2}

\usepackage{dcolumn}
\usepackage{bm}
\usepackage{graphicx,color}
\usepackage[english]{babel}
\usepackage[colorlinks]{hyperref}
\usepackage{comment}
\usepackage{tabularx}
\usepackage{array}
\usepackage[svgnames,table]{xcolor}
\usepackage{ragged2e}

\newcolumntype{A}{>{\centering\arraybackslash \columncolor{white!50!white}}m{2.1cm}}
\newcolumntype{B}{>{\centering\arraybackslash \columncolor{white}}m{7.9cm}}
\newcolumntype{C}{>{\centering\arraybackslash \columncolor{white!50}}m{7.9cm}}
\newcolumntype{D}{>{\centering\arraybackslash \columncolor{white!42}}m}
\newcolumntype{P}[1]{>{\centering\arraybackslash}p{#1}}
\def\beq{\begin{equation}}
\def\eeq{\end{equation}}
\def\bea{\begin{eqnarray}}
\def\eea{\end{eqnarray}}
\def\barr{\begin{array}}
\def\earr{\end{array}}

\usepackage{caption}
\usepackage{color}
\usepackage[font=Large]{subfig}
\usepackage{tabularx,ragged2e,booktabs,caption}
\usepackage{grffile}

\def\ba{\begin{eqnarray}}
\def\ea{\end{eqnarray}}

\begin{document}

\preprint{APS/123-QED}

\title{Variable range hopping in a non-equilibrium steady state}

\author{Preeti Bhandari}
\email{preetibhandari15@gmail.com}
 \affiliation{Department of Physics, Ben-Gurion University of the Negev, Beer Sheva 84105, Israel.}
\author{Vikas Malik}
\email{vikasm76@gmail.com}
 \affiliation{Department of Physics and Material Science, Jaypee Institute of Information Technology, Uttar Pradesh 201309, India.}
\author{Moshe Schechter}%
 \email{smoshe@bgu.ac.il}
\affiliation{
 Department of Physics, Ben-Gurion University of the Negev, Beer Sheva 84105, Israel
}%

\date{\today}

\begin{abstract}
 We propose a Monte Carlo simulation to understand electron transport in a non-equilibrium steady state (\textit{NESS}) for the lattice Coulomb Glass model, created by continuous excitation of single electrons to high energies followed by relaxation of the system. Around the Fermi level, the \textit{NESS} state approximately obeys the Fermi-Dirac statistics, with an effective temperature ($T_{eff}$) greater than the system's bath temperature ($T$). $T_{eff}$ is a function of $T$ and the rate of photon absorption by the system. Furthermore, we find that the change in conductivity is only a function of relaxation times and is almost independent of the bath temperature. Our results indicate that the conductivity of the \textit{NESS} state can still be characterized by the Efros-Shklovskii law with an effective temperature $T_{eff}>T$. Additionally, the dominance of phonon-less hopping over phonon-assisted hopping is used to explain the relevance of the hot-electron model to the conductivity of the \textit{NESS} state.
\end{abstract}

\maketitle

\section{Introduction}
\label{intro}
 The conventional charge transport mechanism at low temperatures in insulators is based on phonon-assisted hopping between localized states. This transport phenomenon is termed variable-range hopping (VRH) \cite{pollak1970effect,shklovskii2013electronic,pollak2013electron}  and is characterized by the conductivity of the form

\begin{equation}
\label{VRH_law}
    \sigma \propto exp \bigg[-\bigg( \frac{T_{0}}{T}\bigg)^{y} \bigg] ,
\end{equation}
where $T$ denotes temperature and $T_0$ denotes a characteristic temperature. For a two-dimensional (2D) system with a constant density of states (DOS), one would expect to see Mott's law of conductivity, i.e., $y = 1/3 $. Efros and Shklovskii \cite{efros1975coulomb} have argued that, due to the presence of Coulomb interactions, the DOS has a so-called Coulomb gap around the Fermi level and follows the relation

\begin{equation}
\label{DOS_law}
   g(\varepsilon) \propto |\varepsilon|^{d-1}   ,
\end{equation}
where $d$ is the spatial dimension and $\varepsilon$, is the Hartree energy. Efros and Shklovskii (ES) further argued that for an interacting system, the exponent $y$ in Eq.(\ref{VRH_law}) is equal to $1/2$ for both two and three dimensions. The conductivity dominated by the VRH mechanism has been observed experimentally \cite{pollak2013electron} and numerically \cite{tsigankov2003long,caravaca2010nonlinear}.

Some interesting physics has come up when an electric field (F) is applied to the system. Depending on the applied field's strength, the conductivity can be in an Ohmic or a non-Ohmic regime. When the applied field is small (low electric field regime), we expect the conductivity to be approximately independent of $F$. This comes under the category of `` Ohmic regime". In this regime, the hopping of electrons is still phonon-assisted (.i.e., depends on the bath temperature only). However, at a high enough applied field, energy cannot be fully dissipated to the phonons, and then the system reaches the ``non-ohmic regime".

Within the non-ohmic regime, for a moderate electric field ($F < kT/e\xi$), the conductivity depends on the field, the temperature, and the localization length ($\xi$)  \cite{hill1971hopping,apsley1975temperature,pollak1976percolation,shklovskii1976nonohmic,van1981hopping}. For systems with small localization length, the conductivity in this regime can be given by the following relation

\begin{equation}
    \label{field_effect}
    \sigma(T,F) = \sigma(T,0) exp \bigg( \frac{eFl_{0}}{kT}\bigg) ,
\end{equation}
 where $l_{0}$ is the typical hopping length and $k$ is the Boltzmann constant. The models satisfying Eq.(\ref{field_effect}) are called Field effect models, \cite{ionov1987nonohmic,timchenko1989variable,grannan1992non,zhang1998non} where the non-linearity in the conductivity is associated with the field-dependent tilt of the energy landscape of electron hopping.
 
 However, systems with large localization lengths are not well characterized by Eq.(\ref{field_effect}), but rather by what is denoted by the ``hot electron model (HEM) \cite{wang1990electrical}. Within the HEM, it is assumed that the conductivity of the system at some temperature (T) and electric field (F) is equal to the linear conductivity at an effective temperature ($T_{eff}$),

\begin{equation}
    \label{HEM}
    \sigma(T,F) = \sigma(T_{eff},0) = \sigma_{0} \, exp \bigg[ -\bigg(\frac{T_{0}}{T_{eff}}\bigg)^{1/2} \bigg] \, .
\end{equation}
Eq.(\ref{HEM}) represents the so-called Hot-electron Model, where $\sigma_{0}$ is the proportionality constant and $T_{0}$ is the characteristic temperature of the system when the ES law is obeyed in the Ohmic regime at temperature T. The assumption here is that the conductivity of the system for a given electric field and temperature depends only on the effective temperature of the electrons ($T_{eff}$) and not on the bath temperature ($T$). Several experiments \cite{leturcq2003hot,galeazzi2007hot,jain2008hot,ladieu2000non} and numerical simulations \cite{caravaca2010nonlinear} on systems that in equilibrium obey VRH have been interpreted in terms of the HEM. 

One can see an activationless hopping when the electric field increases even further, $F > kT/e\xi$ (strong electric field regime) \cite{shklovskii1973hopping,pollak1976percolation,dvurechenski1988activationless,tremblay1989activationless,shahar1990dimensional,yu2004study,kinkhabwala2006numerical,kinkhabwala2006bnumerical}. In this regime, the field plays a role similar to that played by the bath temperature in the ``Ohmic regime", and the conductivity is given by
 
 \begin{equation}
    \sigma \propto exp \bigg[-\bigg( \frac{F_{0}}{F}\bigg)^{1/2} \bigg] \ ,
\end{equation}
  where $F_{0} = b k T_{0}/(e \xi)$. Here $T_{0}$ is the characteristic temperature of the ES law, and $b$ is a constant of order unity. 
 
 The variation in the applied electric field has attracted much attention as it is the formal way of formulating the non-linearity in the conductivity. This method has provided much helpful and reliable information, as discussed before. The present work presents an alternative approach, obtained by creating a non-equilibrium steady state (NESS) by irradiating the sample continuously with high-frequency photons. The conductivity of the NESS state is calculated in the Ohmic regime. The conceptual advantage of this approach is that it decouples the formation of a non-equilibrium state from conductivity calculations. 

 In this paper, we study the Coulomb glass model (see details in Sec. \ref{Hamil}), with localization length twice the lattice spacing, and the equilibrium conductance is in agreement with the ES law. Following the HEM, for each NESS state, dictated by the temperature and the excitation rate, we calculate the effective temperature of the system ($T_{eff}$) related to the occupation of the single electron states near the Fermi level \cite{somoza2008effective}. We then analyze the conductivity ($\sigma$) and density of states at the Fermi level ($g(0)$) through their dependence on $T_{eff}$. Our results are in agreement with HEM results for the field-driven NESS \cite{caravaca2010nonlinear}. For large enough $T_{eff}$, we find deviations from the form of Eq.({\ref{HEM}}). We explain these deviations within the picture of the HEM, their appearance signaling the initiation of the regime where phonon-less conductivity becomes dominant.

 As a second approach, and in an attempt to relate to experimental protocol \cite{ovadyahu2022interaction}, we also analyze the various observables by fitting the equilibrium equations for each observable directly, assigning the temperature as a free parameter. We find that different effective temperatures can be assigned to each observable within this approach. Specifically, we find that the effective temperature for the conductivity is smaller than the effective temperature for the density of states at the Fermi energy. This result is in qualitative agreement with a recent experiment on Indium oxide films where the NESS state under the influence of IR radiation was studied. However, the experimental results show larger deviations between the effective temperatures of the conductivity and of the memory dip; see the discussion below.
 
   Our paper is organized as follows. In Sec. {\ref{Hamil}}, the Coulomb Glass lattice model is introduced. We present the numerical techniques used here in Sec. {\ref{numerics}}. In Sec. {\ref{Result}}, we present our numerical results, and finally, in Sec. {\ref{Discuss}}, we conclude the paper with a summary of our results.
 
\begin{figure}[t]
\centering
\includegraphics[scale=0.55]{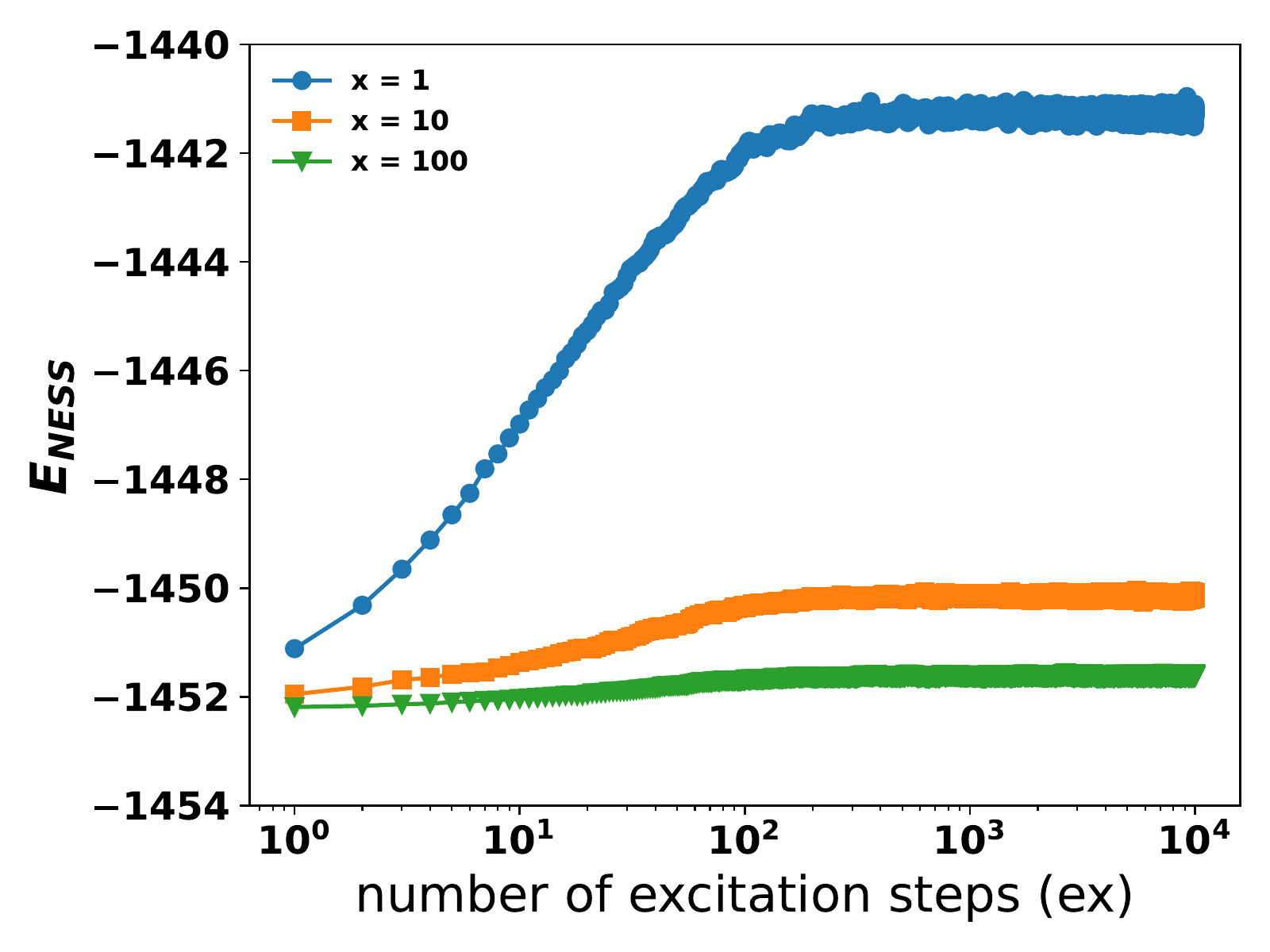}
\caption{\label{ENESS_state} Energy of the non-equilibrium steady state as a function of the number of excitation steps at different relaxation steps. Different graphs correspond to a different number of relaxation steps ($x$) following each excitation step. The initial state here is the annealed state at $\beta=80$.}.  
\end{figure}

\section{Model}
\label{Hamil}
We consider the standard two-dimensional (2d) Coulomb Glass (CG) lattice model with the Hamiltonian

\begin{equation}
    \label{Hamiltonian}
    H = \sum_{i} \phi_{i} S_{i} + \frac{1}{2} \sum_{i \neq j} \frac{S_{i} S_{j}}{r_{ij}} ,
\end{equation}
where $S_{i} = n_{i} - 1/2$ is the pseudo spin variable at site $i$, $n_{i} \in \{ 0,1 \}$ is the electron occupation number, $\phi_{i}$ are the random on-site energies chosen from a box distribution with the interval $\bigg[ -W/2, W/2\bigg]$ and $r_{ij}$ is the distance between sites $i$ and $j$ under periodic boundary conditions. All energies and temperatures are calculated in units of $e^2/a$, where $a$ is the lattice constant. We have used a square lattice of size $64 \times 64$ and disorder $W = 2$. 

\section{Numerical details}
\label{numerics}
 Experimentally, the NESS state is created by irradiating a well-equilibrated sample with high-frequency radiation ($\nu$), having energy greater than the Coulomb gap width ($\delta$), $h\nu$ $>> \delta$. Numerically, to create a NESS state, we first use the simulated annealing technique \cite{bhandari2019finite} to thermalize the Hamiltonian (\ref{Hamiltonian}). All the observables were averaged over time (i.e., logarithmic binning of Monte Carlo steps) to test the equilibration. The system is in thermal equilibrium once the last three bins agree within errors. A two-step approach is required to create a NESS condition at a specific temperature. The first step involves inducing excitation in the system (i.e., $\Delta E = \varepsilon_{i} - \varepsilon_{j} - 1/r_{ij} > 0$), followed by a relaxation procedure. These two steps are then repeated continuously.

\textit{Excitation process:} We start with an annealed state (we have considered the annealed states at $\beta=1/T=20,40$ and $80$) and choose a site `$i$' at random. Then we pick another site `$j$', now with a probability, $e^{-2 r_{ij}/\xi}$, where $\xi=2$ is the localization length. In addition, we make certain that $S_{i} \neq S_{j}$. Finally, we swap the two spins, creating an excitation in the system. This corresponds to an electron making a $\Delta E > 0$ transition between site $i$ and $j$ by absorbing a photon.

\textit{Relaxation process:} Experimentally, the time between two photons striking the system (relaxation time, $x$) varies as a function of the power. 
\begin{equation}
    \label{power}
    Power \propto \frac{1}{relaxation \, time} \quad .
\end{equation}
Between two-photon absorptions, the system may relax toward the equilibrium state. To simulate the relaxation process, we allow the system to relax via the kinetic monte carlo algorithm \cite{tsigankov2003long} between two subsequent excitations. In our simulation, we vary the number of relaxation steps ($x = 1, 3, 7, 10, 30,70, 100$) after each excitation process ($ex$) and trace its effect on the energy, DOS, occupation probabilities, and conductivity of the NESS state. A single relaxation step corresponds to $L^{2}$ Monte Carlo steps.

A NESS state is reached once the rate of energy gained by the excitations becomes equal to the energy lost in the relaxation process, and the system achieves steady-state energy. The energy of the NESS state increases as the relaxation time decreases, as seen in Fig.(\ref{ENESS_state}). 

To simulate the conductivity of the NESS state, we apply a small electric field (F) in the x-direction by adding a term $\sum_{i} F x_{i}$ (where $F=T/10$) to the Hamiltonian (\ref{Hamiltonian}), and perform Kinetic Monte Carlo simulations as described in Ref.\cite{tsigankov2003long} The NESS state is maintained during conductivity calculations, and results were obtained after averaging over 500 disorder realizations.

\begin{figure*}[t]
\centering
\includegraphics[scale=0.5]{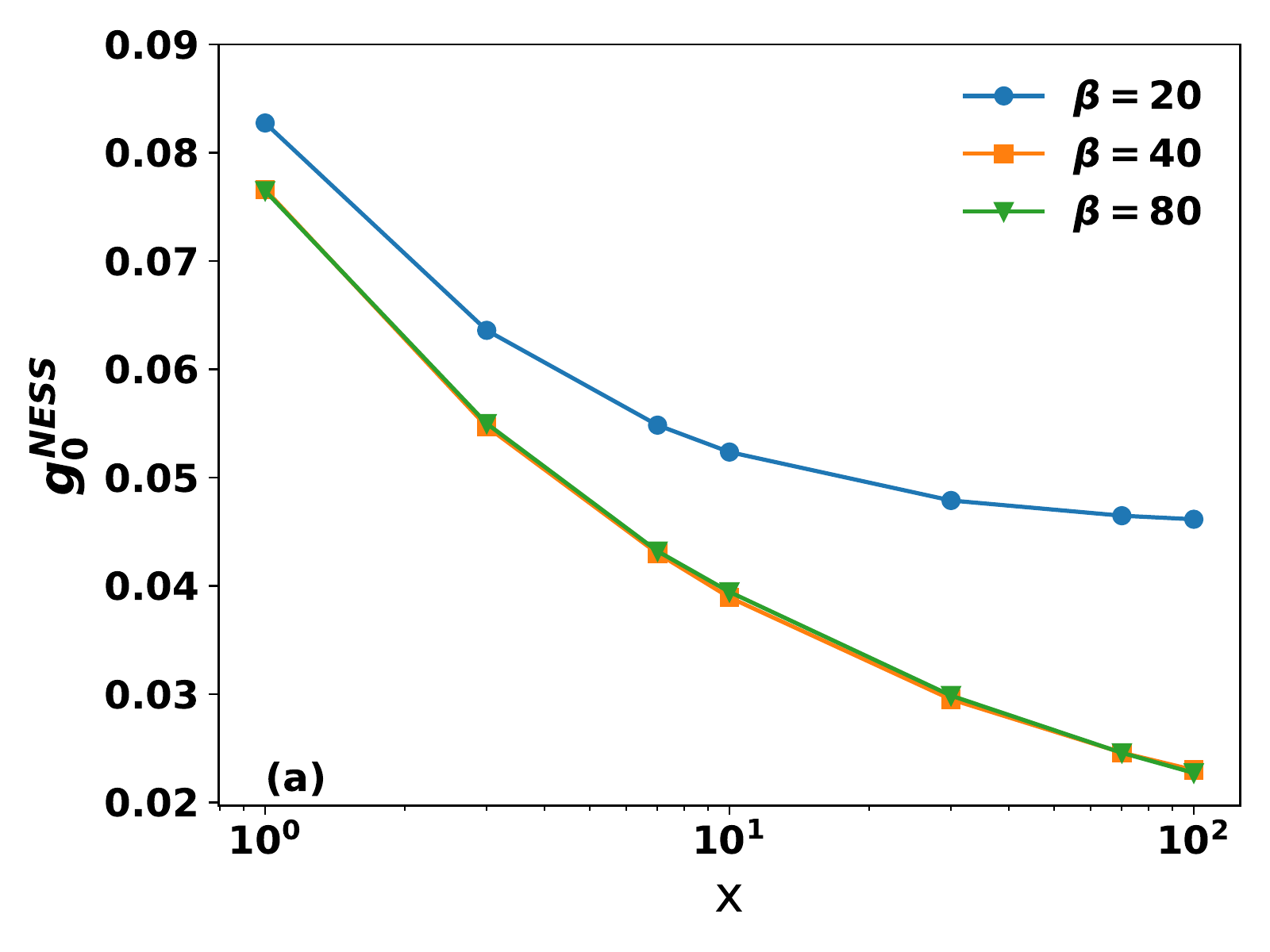}
\includegraphics[scale=0.5]{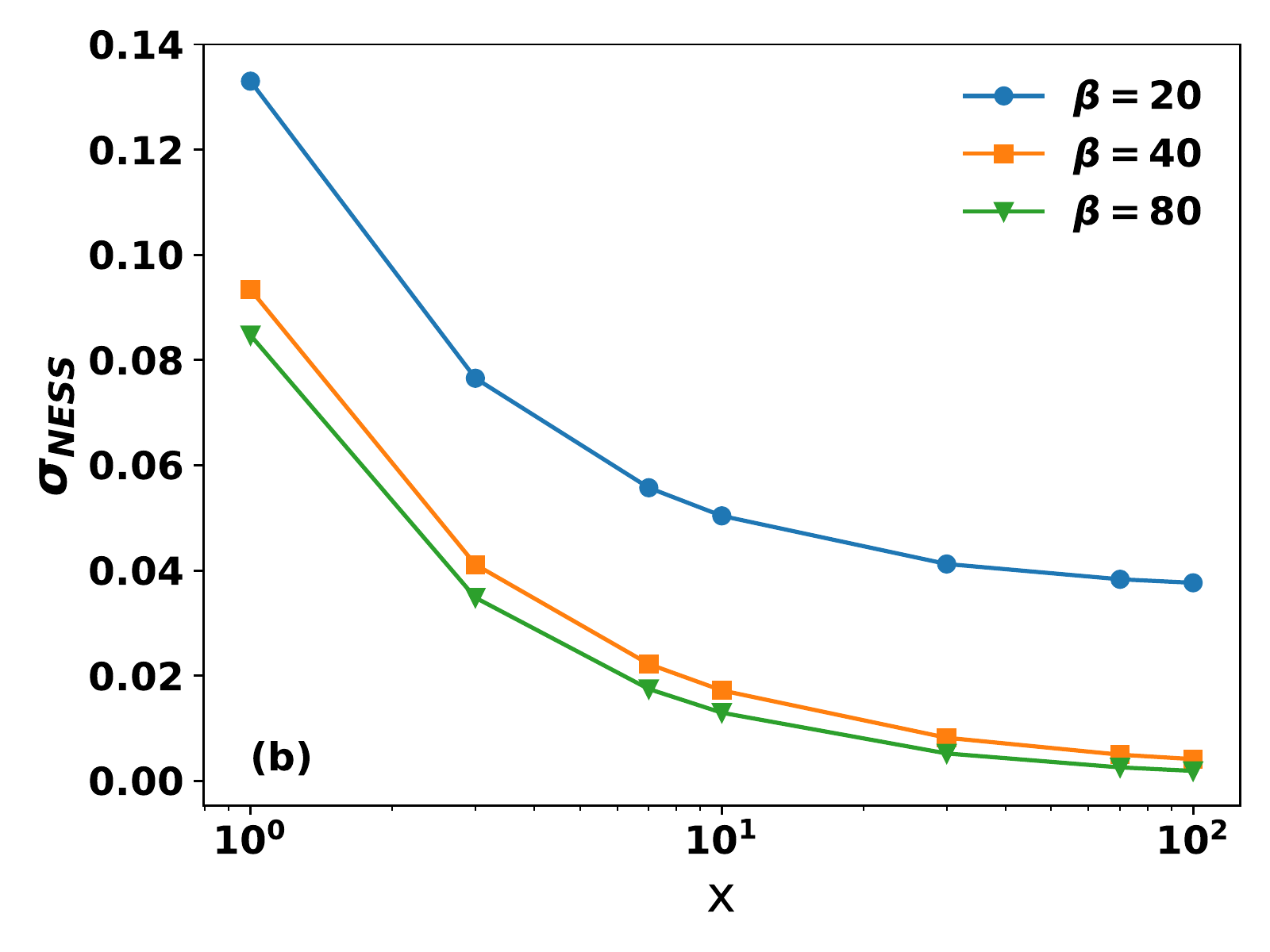}
\caption{\label{order_paametes} (a) The behavior of the density of states at the Fermi level (for the NESS state) as a function of relaxation steps ($x$) at different bath temperatures ($T = 1/\beta$). (b) The conductivity of the NESS state as a function of $x$ at different bath temperatures. }
\end{figure*}
 
\section{Results}
\label{Result} 

The perturbation leading to the NESS state drives the system out of equilibrium. Specifically, the relaxation of the electrons near the Fermi energy is limited by the scarcity of available states to relax to. This leads to a change of the occupation probability of electronic states near the Fermi energy \cite{somoza2008effective}, which is well approximated by a Fermi function with $T_{eff} \neq T$,
\begin{equation}
\label{FD_Teff}
f_{i} = \frac{1}{[exp(\varepsilon_{i}/T_{eff}) + 1]} \quad .    \end{equation}
The supplemental material contains more information on calculating $T_{eff}$ using Eq.(\ref{FD_Teff}). This effective temperature is used to express the various quantities below.

Figure {\ref{order_paametes}} shows the behavior of the DOS at the Fermi-level ($g(0)_{NESS}$) and the conductivity ($\sigma_{NESS}$) of the NESS state as a function of $x$ at different temperatures. 

\begin{figure}[b]
\centering
\includegraphics[scale=0.55]{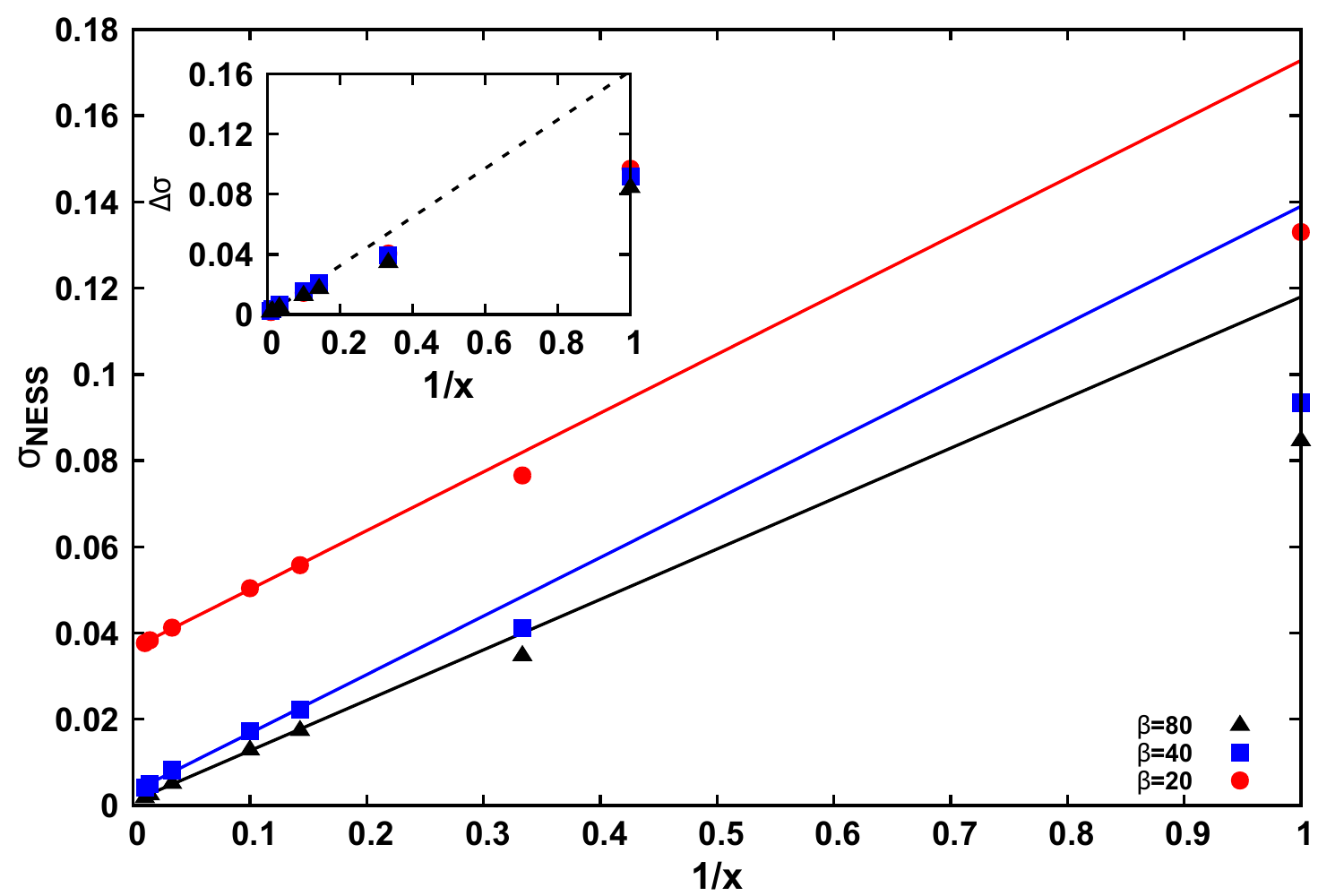}
\caption{\label{del_sigma_1_relax} The conductivity of the NESS state at various bath temperatures as a function of intensity $(1/x)$. The behavior of change in conductivity as a function of intensity for the NESS state at different bath temperatures is presented in the inset.}
\end{figure}

Let us first consider the NESS state's conductivity in more detail. In Fig.(\ref{del_sigma_1_relax}), we plot $\sigma_{NESS}$ as a function of $x$. At long relaxation times, we find 
\begin{equation}
    \sigma_{NESS} \approx \sigma(T) + \frac{c}{x} \ ,
\end{equation}
.i.e. $\sigma_{NESS} - \sigma(T)$ is nearly independent of $T$, as shown in the inset. Here $c$ is a constant, and $\sigma(T)$ is the conductivity of the thermal state.

\begin{figure}[b]
	\includegraphics[scale=0.55]{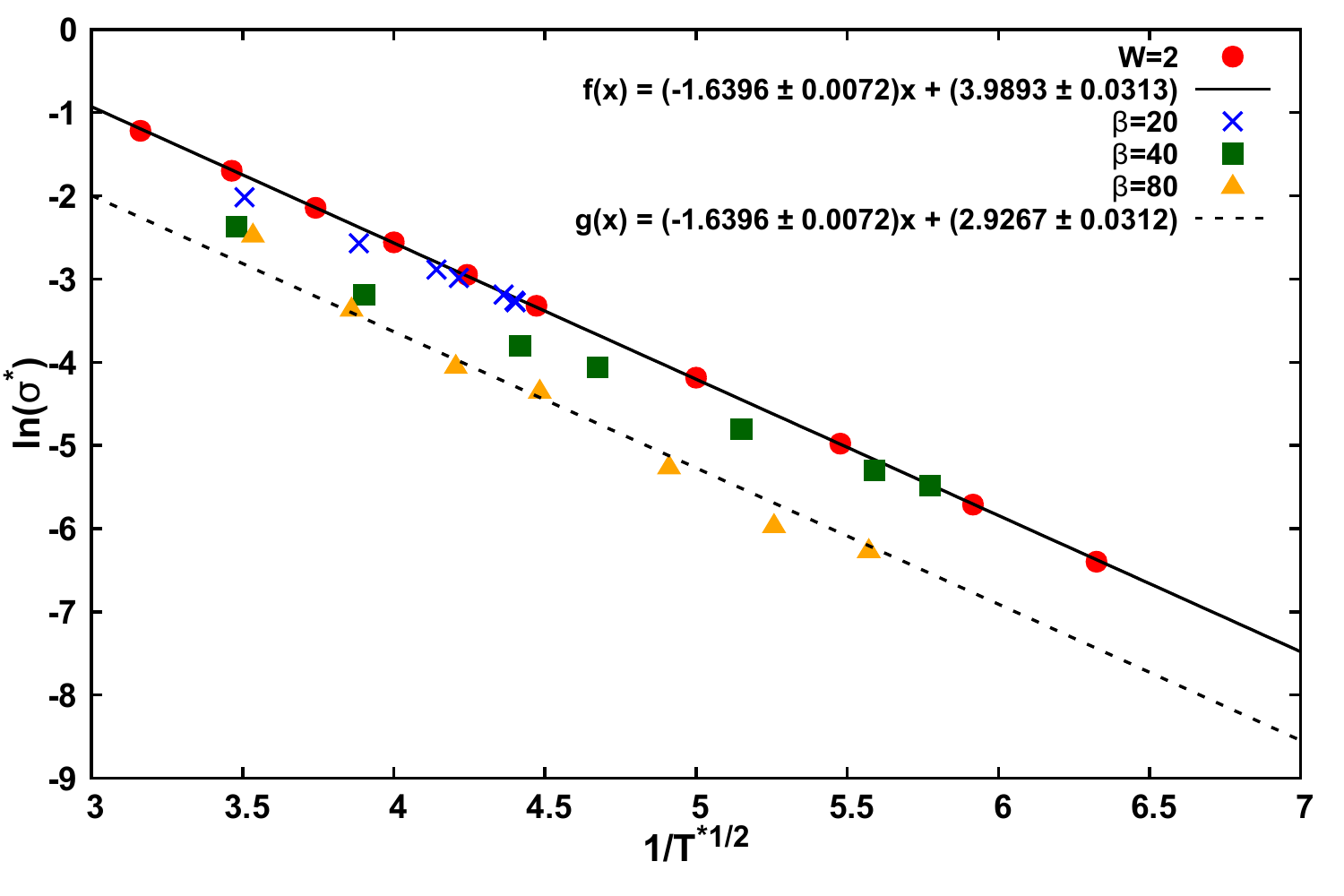}
	\caption{\label{T_half_law} Conductivity ($\sigma^{*}$) as a function of temperature ($T^{*}$) is log-real plot. Here $\sigma^{*} = \sigma$ and $T^{*}=T$ for the equilibrium data and $\sigma^{*} = \sigma_{NESS}$ and $T^{*}=T_{eff}$ for the NESS data. The red dots correspond to the conductivity of the thermal state. This data has been fitted using  the relation, $\sigma = \sigma_{0} \ exp\{-(T_{0}/T)^{1/2}\}$ (solid line). Blue cross ($\beta=20$), green square ($\beta=40$), and orange triangle ($\beta=80$) correspond to the conductivity of the NESS state as a function of $T_{eff}^{-1/2}$. The orange triangles are fitted (dashed line) using the relation $\sigma_{ness} = \sigma^{\prime}_{0} \ exp\{-(T_{0}/T_{eff})^{1/2}\}$, where $T_{eff}$ is calculated from the Fermi Dirac distribution.}
\end{figure}

We now analyze our results for the conductivity with regard to the Hot Electron Model. Our analysis of the HEM is presented in Fig.(\ref{T_half_law}), where we have used $T_{eff}$ computed using Eq.(\ref{FD_Teff}) and plotted $\sigma_{NESS}$ vs $T_{eff}^{-1/2}$ for different temperatures and relaxation times. The solid line in Fig.(\ref{T_half_law}) corresponds to the conductivity as given by Eq.(\ref{HEM}).

For $\beta=20$ and small enough intensity (large relaxation times $x > 7$), Eq.(\ref{HEM}) is approximately satisfied. For $\beta=40$ and $80$, the deviation from the solid line in Fig.(\ref{T_half_law}) leads us to a conclusion that for low temperatures (and for cases where $\Delta \sigma >> \sigma$), the conductivity of the NESS state can be explained using the following relation

\begin{equation}
    \label{HEM_new}
    \sigma_{NESS} = \sigma_{0}^{\prime} \, exp \bigg[ -\bigg(\frac{T_{0}^{\prime}}{T_{eff}} \bigg)^{1/2} \bigg] \ ,
\end{equation}
where $T_{0}^{\prime} \approx T_{0}$ and $\sigma_{0}^{\prime} \geq \sigma_{0}/2$ approaching $\sigma_{0}/2$ for $T_{eff} \textgreater\textgreater T$ \cite{pollak2013electron}. This variation of $\sigma_{0}$ from the constant value given in Eq.(\ref{HEM}) provides a generalization of the HEM. Let us now discuss its physical origin. In the case of a well-equilibrated thermal state, up transitions ($\Delta E > 0$) and down transitions ($\Delta E < 0$) contribute equally to the conductivity of the system, and the transition from a site $i$ to $j$ is given by

\begin{equation}
    \label{rate_eqn}
    \Gamma_{ij} \sim \gamma_{0} \, exp \bigg[ -\frac{2r_{ij}}{\xi} - \frac{\Delta E_{ij}}{kT} \bigg] \quad {\color{red}.}
\end{equation}
Minimizing the exponent in Eq.(\ref{rate_eqn}) leads to the ES law of conductivity (Eq.(\ref{VRH_law})). Note that $\Gamma_{ij} = \Gamma_{ji}$ for a system at a steady state under the influence of a small electric field. On the limited range of temperatures available in our simulation, the temperature dependence of the conductivity in equilibrium is in agreement with the ES law (as shown by the solid line in Fig.(\ref{T_half_law}).

Now for a NESS state, where the change in conductivity is mainly due to the downward energy transitions, $\Gamma_{ij} \neq \Gamma_{ji}$. Here, the downward energy transitions between two sites, $i$ and $j$, given by 
\begin{equation}
    \label{rate_eqn2}
    \Gamma_{ij} = \gamma_{0} \, exp \bigg[ -\frac{2r_{ij}}{\xi} - \frac{\Delta E_{ij}}{k \, T_{eff}} \bigg]. 
\end{equation}
The reverse upward energy transition from $j$ to $i$ is given by Eq.(\ref{rate_eqn}). When $T_{eff} >> T$, the energy-lowering transitions dominate over energy-gaining ones, and the system's conductivity is defined by Eq.(\ref{HEM_new}). When the downward transitions completely dominate the system (we see this at very low temperature $\beta=80$), then

\begin{equation}
   \label{our_eqn}
    \sigma_{NESS}(T_{eff},x) \approx \frac{1}{2} \sigma(T_{eff}) \ .
\end{equation}

In Fig.(\ref{T_half_law}), the deviation of the NESS conductivity (dashed line) from the equilibrium conductivity (solid line) is a feature of the HEM and not its failure. The change in Fig.(\ref{T_half_law}) from $\sigma_{0}$ being similar to its thermal value to being about half its thermal value notes the relaxation time (value of $x$ in our case) where the phonon-less conductivity becomes dominant. We note that for each temperature, there is a one-to-one correspondence between the effective temperature and the relaxation time ($x$).

\begin{figure}
\centering
\includegraphics[scale=0.55]{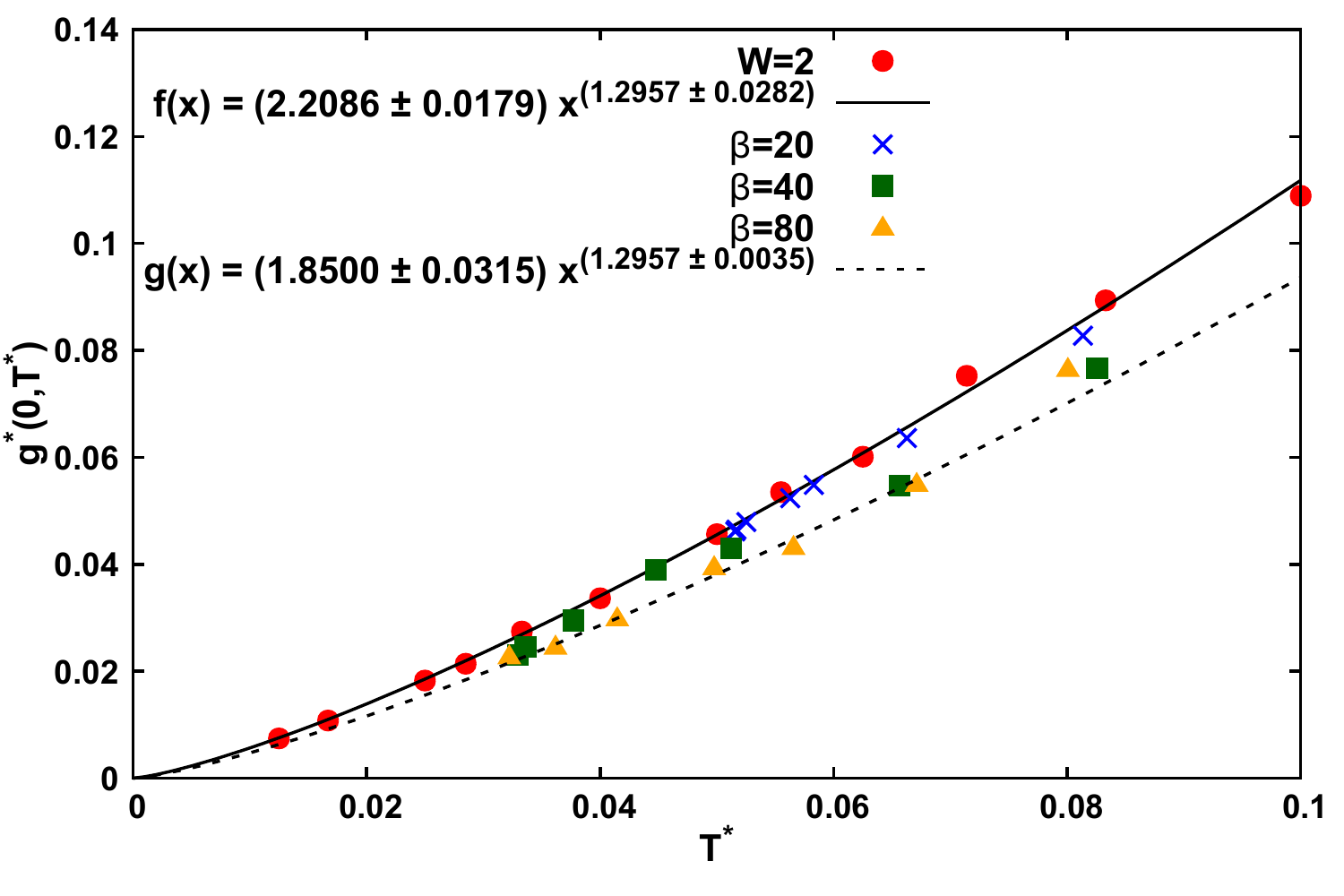}
\caption{\label{g_Teff} Density of state at the Fermi-level ($g^{*}(0,T^{*})$) as a function of temperature ($T^{*}$). Here $g^{*}(0,T^{*}) = g(0,T)$ and $T^{*}=T$ for the equilibrium data and $g^{*}(0,T^{*}) = g^{NESS}(0,T_{eff})$ and $T^{*}=T_{eff}$ for the NESS data. The red dots correspond to the single-particle density of states (DOS) at the Fermi-level of the thermal states $g(0, T)$ at disorder $W=2$. This data has been fitted using  the relation, $g(0,T) = c \, T^{\alpha}$ (solid line). Blue cross ($\beta=20$), green square ($\beta=40$), and orange triangle ($\beta=80$) correspond to DOS of the NESS state as a function of $T_{eff}$. The orange triangles are fitted (dashed line) using the relation $g^{NESS}(0, T_{eff}) = c^{\prime} \, T_{eff}^{\alpha}$, where $T_{eff}$ is calculated from the Fermi Dirac distribution. }
\end{figure}

We can now also explain the results shown in Fig.(\ref{del_sigma_1_relax}). The system gains excess energy (excitations) by photon absorption and attempts to attain equilibrium by energy-lowering transitions. The number of excitations created by photon absorption is proportional to the radiation power. Thus, the excess conductivity, i.e., $\Delta{\sigma}$, dominated by energy-lowering transitions facilitated by excited electrons, depends only on the radiation power, not on the bath temperature (see the supplementary material for more details). 

Another experimentally relevant quantity is the DOS at the Fermi-level $g(0)$, which, theoretically for 2D systems, is proportional to the phonon-bath temperature (T) \cite{mogilyanskii1989self}. Numerical simulations claims that $g(0,T) \propto T^{\alpha}$ where $\alpha \neq 1$ \cite{mobius1992coulomb,sarvestani1995coulomb}. As shown in Fig.(\ref{g_Teff}) (red circles), our data show that the density of states of the thermal states at the Fermi-level $g(0)$ follows the following relation
\begin{equation}
    \label{dos_thermal}
    g(0) =c T^{\alpha} \ , 
\end{equation}
where $\alpha=1.29$ and $c$ is the proportionality constant. For the NESS state, we find (see Fig.(\ref{g_Teff})) that the density of states at the Fermi-level, $g_{0}^{NESS}(x, T)$, as a function of relaxation time $(x)$ and bath temperature $T$ follows the relation
\begin{equation}
    \label{dos_ness}
    g_{0}^{NESS}(x,T) = c^{\prime} (T_{eff})^{\alpha} \ ,
\end{equation}
 where $c \approx c^{\prime}$ at $\beta = 20$ and $c \neq c^{\prime}$ as the temperature and the relaxation time decreases. Also, here, we find a decrease in the proportionality constant in the regime where phonon-less hopping dominates, albeit this effect here is smaller than for the conductivity.
 
\begin{figure}[t]
\centering
\includegraphics[scale=0.55]{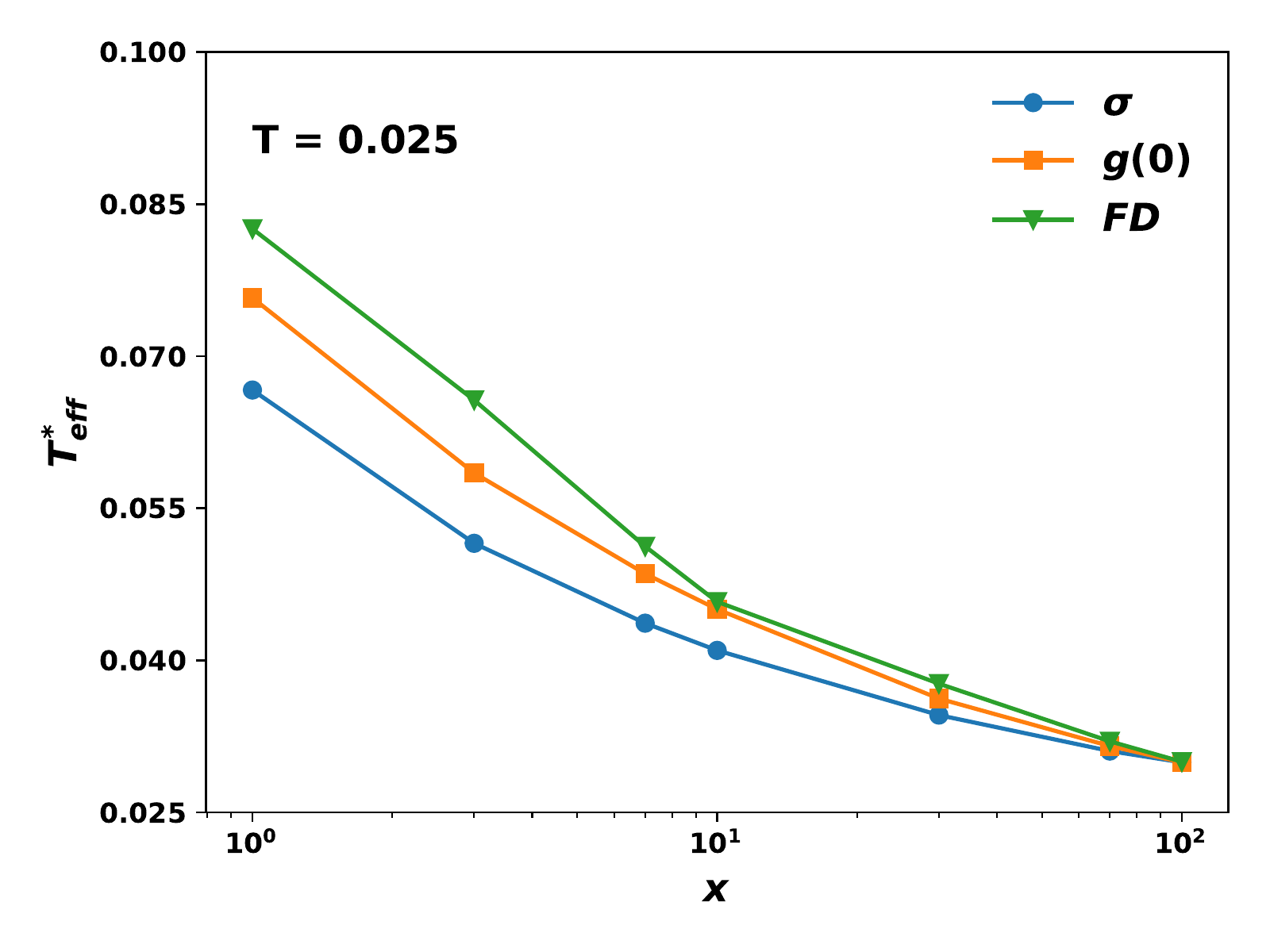}
\caption{\label{NESS_b80_Teff} Effective temperature ($T_{eff}^{*}$) of the NESS state as a function of relaxation time ($x$) calculated using: conductivity ($\sigma$), Fermi-Dirac distribution around the Fermi-level (FD), and DOS at the Fermi-level ($g(0)$)) at the bath temperature $T = 0.025$. $T_{eff}^{*}$ here corresponds to $T_{eff}^{\sigma}$ for the blue line, $T_{eff}^{g}$ for the orange line and $T_{eff}^{FD}$ for the green line. }
\end{figure}

\begin{figure}[t]
\centering
\includegraphics[scale=0.55]{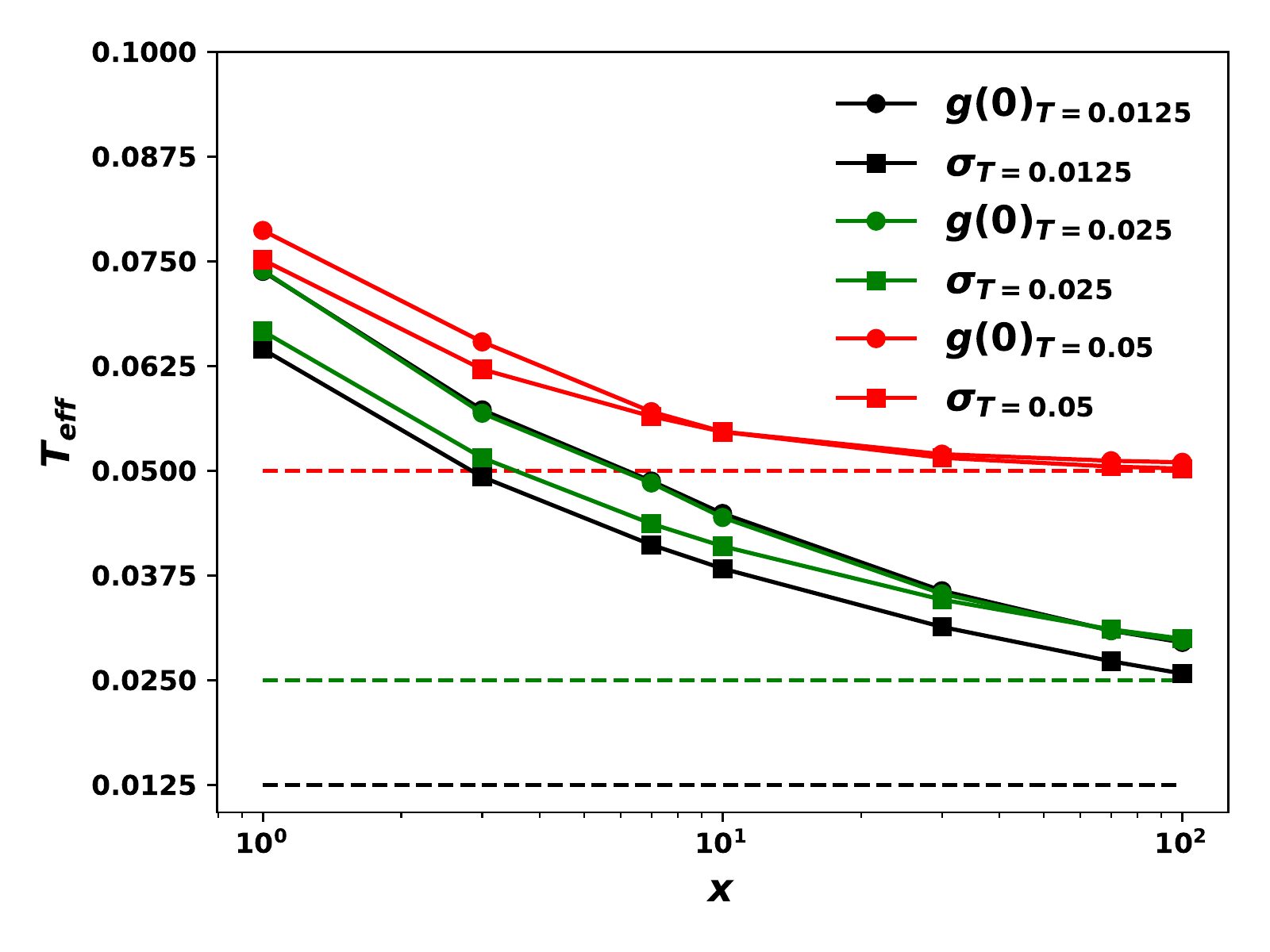}
\caption{\label{T_eff} Effective temperature for a NESS state at beta = 20, 40, and 80 as a function of relaxation times ($x$). $T^{g}_{eff}$ and $T^{\sigma}_{eff}$ are different when the system is out of equilibrium, which gradually merges to a single temperature ($T$) as the relaxation time increases (system approaching equilibrium). The lower the temperature, the longer the relaxation times where the two effective temperatures merge.} 
\end{figure}

Finally, we now make an attempt to connect our numerical simulation to recent experimental results on the NESS state of amorphous indium oxide \cite{ovadyahu2022interaction}. Here, following the analysis in the experiment, we adopt a different method to determine the effective temperature for conductivity ($T^{\sigma}_{eff}$) and DOS at the Fermi level ($T^{g}_{eff}$). (The detailed explanation of the procedure used to calculate the effective temperature is provided in the supplementary material). We calculate $T^{\sigma}_{eff}$ and $T^{g}_{eff}$ of the NESS state by using Eq.(\ref{HEM}) and Eq.(\ref{dos_thermal}) respectively at different temperatures and relaxation times. The values of the parameters $c$, $\sigma_{0}$, $T_{0}$ are kept equal to their equilibrium values in this calculation. For clarity, we denote the $T_{eff}$ calculated using FD distribution in Eq.(\ref{FD_Teff}) as $T^{FD}_{eff}$ for the rest of the discussion. We observe that at short relaxation times, the effective temperatures describing each physical quantity are different. In Fig.(\ref{NESS_b80_Teff}), we plot the effective temperatures for the different observables as a function of relaxation time at $T = 0.025$ (data for $T = 0.05$ and $T = 0.0125$ are shown in supplementary material). Note that there is a threshold degree of non-equilibrium, where $\Delta \sigma > \sigma$ (which depends on the bath temperature and relaxation time of the system) only beyond which effective temperatures of different observables appears. Unlike the situation in equilibrium, when the system is pushed far enough out of equilibrium, it acquires multiple time scales affecting differently various measurable quantities and consequently dictating observable specific effective temperatures.

Let us note that for the regime where the effective temperatures are different, $T^{\sigma}_{eff}$ is always less than $T^{FD}_{eff}$. This can be explained as follows: Comparing the expressions for the conductivity within the two approaches discussed above, we find
\begin{equation}
   \label{sigma_solve}
    \sigma_{0} \, exp \bigg[ -\bigg( \frac{T_{0}}{T_{eff}^{\sigma}} \bigg)^{1/2} \bigg] = \sigma_{0}^{\prime} \, exp \bigg[ -\bigg( \frac{T_{0}}{T^{FD}_{eff}} \bigg)^{1/2} \bigg] \ .
\end{equation}
At low temperatures, one finds that $\sigma_{0}^{\prime} \approx \sigma_{0}/2$  (this is true in our case at $\beta=80$ where the phononless hopping completely dominates). Using this relation in Eq.(\ref{sigma_solve}), we get

\begin{equation}
  \bigg(\frac{1}{T_{eff}^{\sigma}} \bigg)^{1/2} \approx \bigg(\frac{1}{T_{eff}^{FD}} \bigg)^{1/2} + \bigg(\frac{ln(2)}{\sqrt{T_{0}}} \bigg) \ ,
\end{equation}
i.e. $T_{eff}^{\sigma} < T_{eff}^{FD}$.

 We now compare our findings with the experiment \cite{ovadyahu2022interaction}, where it was found that far from equilibrium, the effective temperature of the conductivity is much smaller than the effective temperature of the memory dip. Following this experimental finding, we concentrate on the effective temperatures of the NESS state that we determined using $\sigma$ and $g(0)$; the latter is believed to be proportional to the memory dip. We compare the two effective temperatures at T=0.05, 0.025, and 0.0125 for different relaxation times, as shown in Fig.(\ref{T_eff}). The dashed line represents the bath temperature and indicates how far the NESS state is from the equilibrium state. When the system is further from equilibrium, we do observe that the effective temperature computed using conductivity ($T^{\sigma}_{eff}$) is lower than the one computed using the density of states at the Fermi level ($T^{g}_{eff}$). However, this difference is much smaller than the difference between the experimentally obtained effective temperatures for the conductivity and the DOS. Fig.(\ref{T_eff}) also shows that as the system moves away from equilibrium, the two effective temperatures converge at a larger relaxation time value.

\section{Discussion}
\label{Discuss} We demonstrate the electron transport in a non-equilibrium steady state in the context of understanding the general theory of the Hot-electron model. The non-equilibrium steady state is created by irradiating the system with high-frequency photons. Experiments of this kind have been recently conducted on indium oxide films \cite{ovadyahu2022interaction}.

We have calculated the conductivity ($\sigma$) and the single-particle density of states at the Fermi-level ($g(0)$) of the NESS state. At relatively high temperatures, our results are remarkably similar to the non-equilibrium simulation results by Caravaca \textit{et} \textit{al} \cite{caravaca2010nonlinear} done in high electric fields at $\xi = 2$. At lower temperatures, at first glance (in our work) and as reported by Caravaca \textit{et} \textit{al} \cite{caravaca2010nonlinear} for smaller localization length ($\xi = 1$ case), the conductivity of the NESS state deviates from the HEM. The deviation happens because of slow relaxation due to small $\xi$ in Ref. (\cite{caravaca2010nonlinear}) and small $T$ in our case. We explain this deviation as a feature of the HEM in terms of the dominance in this regime of phonon-less hopping over phonon-assisted hopping. We show that both $g(0)$ and $\sigma$ of the non-equilibrium steady state obey the HEM model.

Our results, therefore, provide a robust way to understand the crossover from phonon-less hopping to phonon-assisted hopping, which can be tested experimentally by varying the intensity of radiation on the target sample or decreasing the temperature of the system.

Our results are also qualitatively in agreement with the experimental finding in \cite{ovadyahu2022interaction} that $T^{\sigma}_{eff}$ $<$ $T^{g}_{eff}$ when the NESS state is far from equilibrium. One of the possible reasons for the quantitative difference could be that the measurement in the experiment (memory dip) is not directly a measurement of the DOS (in simulations) but is believed to be proportional to it. Another possibility can be quantum effects which are beyond the current study in this paper.

In the future, it will be interesting to explore how our results alter if we move from the ES to the Mott regime. It would also be interesting to see how positional disorder affects the system.

\section*{Acknowledgement} 
P.B. acknowledges the Kreitman School of Advanced Graduate Studies for financial support. M.S. acknowledges support from the Israel Science Foundation (Grant No. 2300/19). Illuminating discussions with Z. Ovadyahu is gratefully acknowledged. 

%

\end{document}


\begin{center}
\textbf{Supplemental Material: Variable range hopping in a non-equilibrium steady state}
\end{center}
\vskip 2pt

\section{Calculation of effective temperature}

In this paper, we have calculated the effective temperature using three different procedures. The details of the procedure are as follows:

\underline{\textit{Using Fermi-Dirac distribution}}: Once the system reaches a steady state, we calculate the site occupation probability ($f_{i}$) as a function of Hartree energy ($\varepsilon_{i}$), where

\begin{equation}
    \label{HE}
    \varepsilon_{i} = \phi_{i} + \sum_{j \neq i} \frac{n_{j}}{r_{ij}}
\end{equation}
We find that around the Fermi-level, $f_{i}$ is well approximated by the Fermi-Dirac distribution with an effective temperature, i.e., 
\begin{equation}
\label{FD_Teff}
f_{i} = \frac{1}{[exp(\varepsilon_{i}/T_{eff}) + 1]}.    
\end{equation}

\begin{figure*}[h]
    \centering
    \includegraphics[scale=0.55]{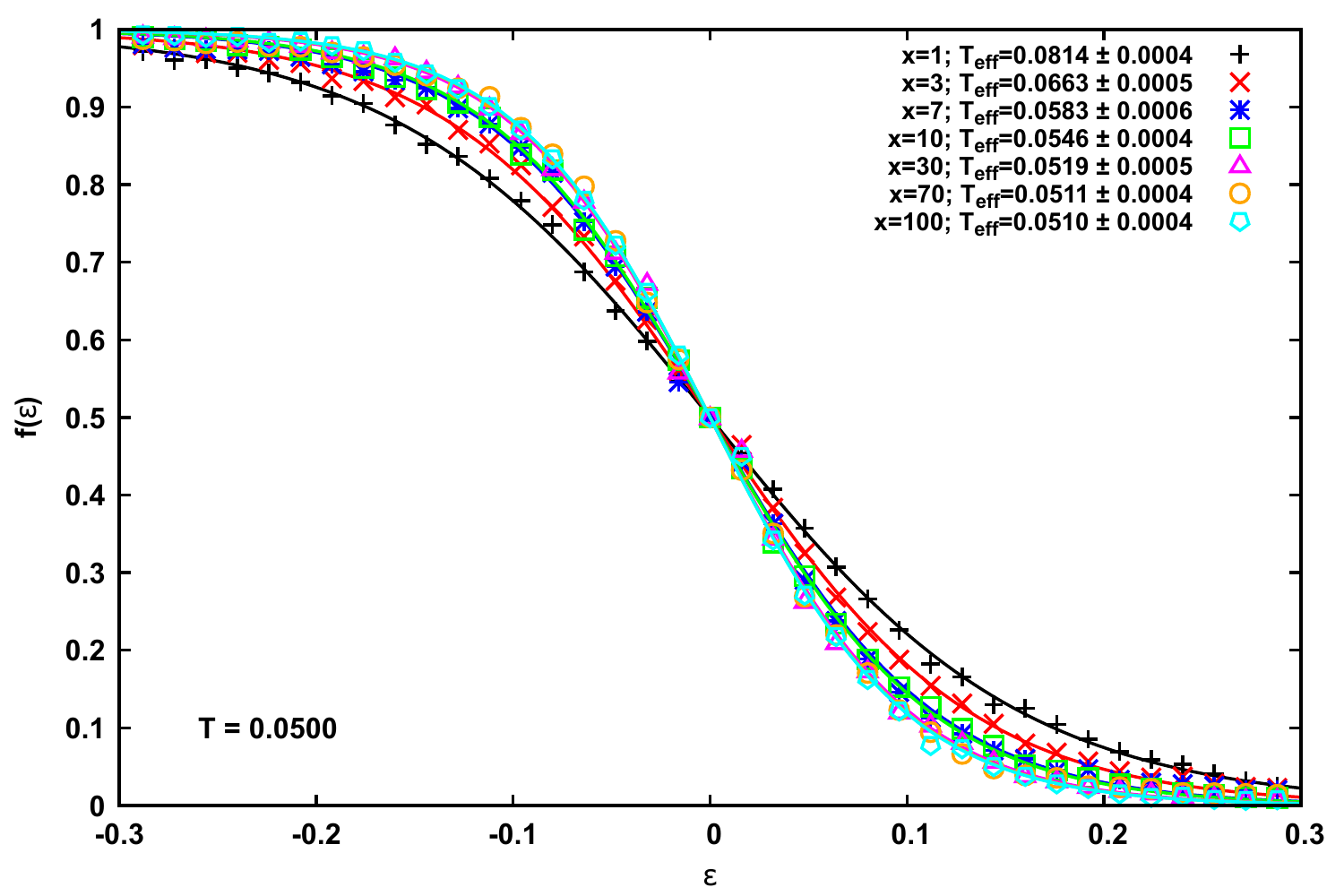}
    \includegraphics[scale=0.55]{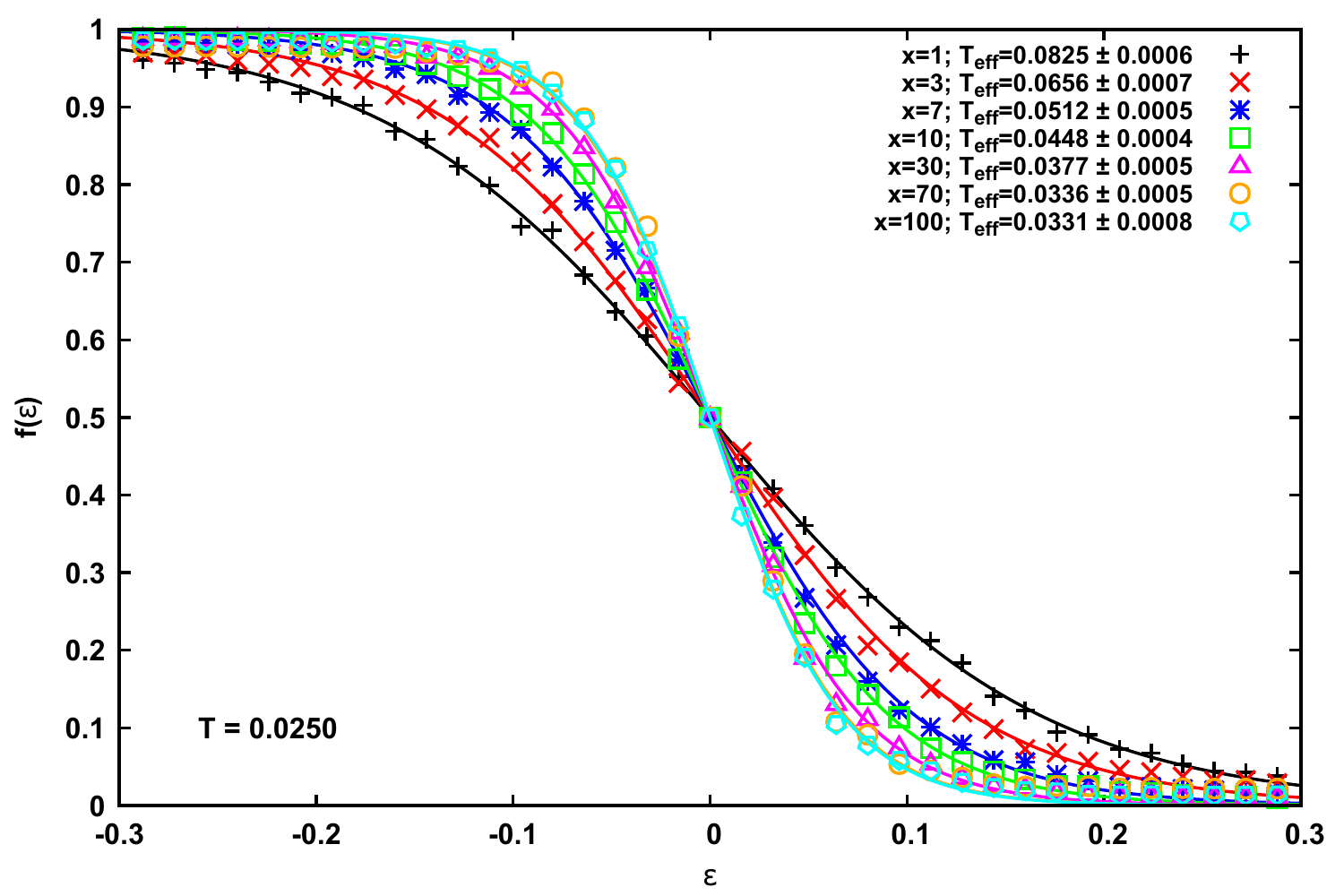}
    \includegraphics[scale=0.55]{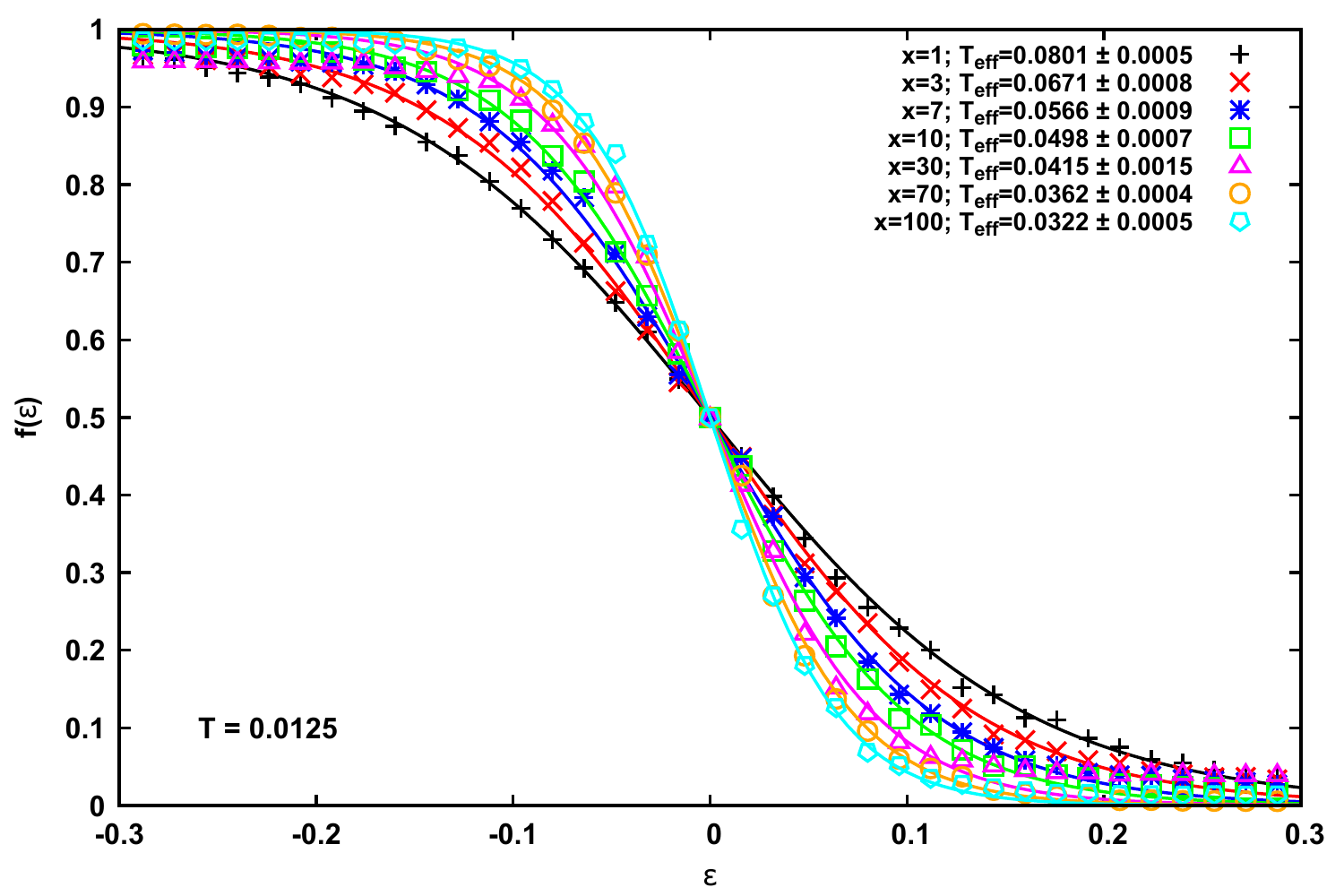}
    \caption{Site occupation probability as a function
of Hartree energy. The continuous curves are fit by Fermi-distribution using Eq.(\ref{FD_Teff}).}
    \label{fig:my_label}
\end{figure*}

\underline{\textit{Using conductivity relation}}: As shown in the main text (red circles in Fig.4 in the main text) and Fig.2 here, our data shows that in the ohmic regime, the ES law is satisfied for the temperature range studied here, 
\begin{equation}
    \label{ES_law}
    \sigma  = \sigma_{0} exp[-(T_{0}/T)^{1/2}].
\end{equation}

In Eq.(\ref{ES_law}), we replace $\sigma$ by $\sigma_{NESS}$ and calculate the effective temperature $T_{eff}^{\sigma}$ of the NESS state as a function of bath temperature $T$ and relaxation time $x$.
\begin{equation}
    \label{sigma_ness}
    \sigma_{NESS}(T,x) = \sigma_{0} \, exp\bigg[ -\bigg(\frac{T_{0}}{T_{eff}^{\sigma}} \bigg)^{1/2}\bigg]
\end{equation}

\begin{figure}[h]
    \centering
    \includegraphics[scale=1.0]{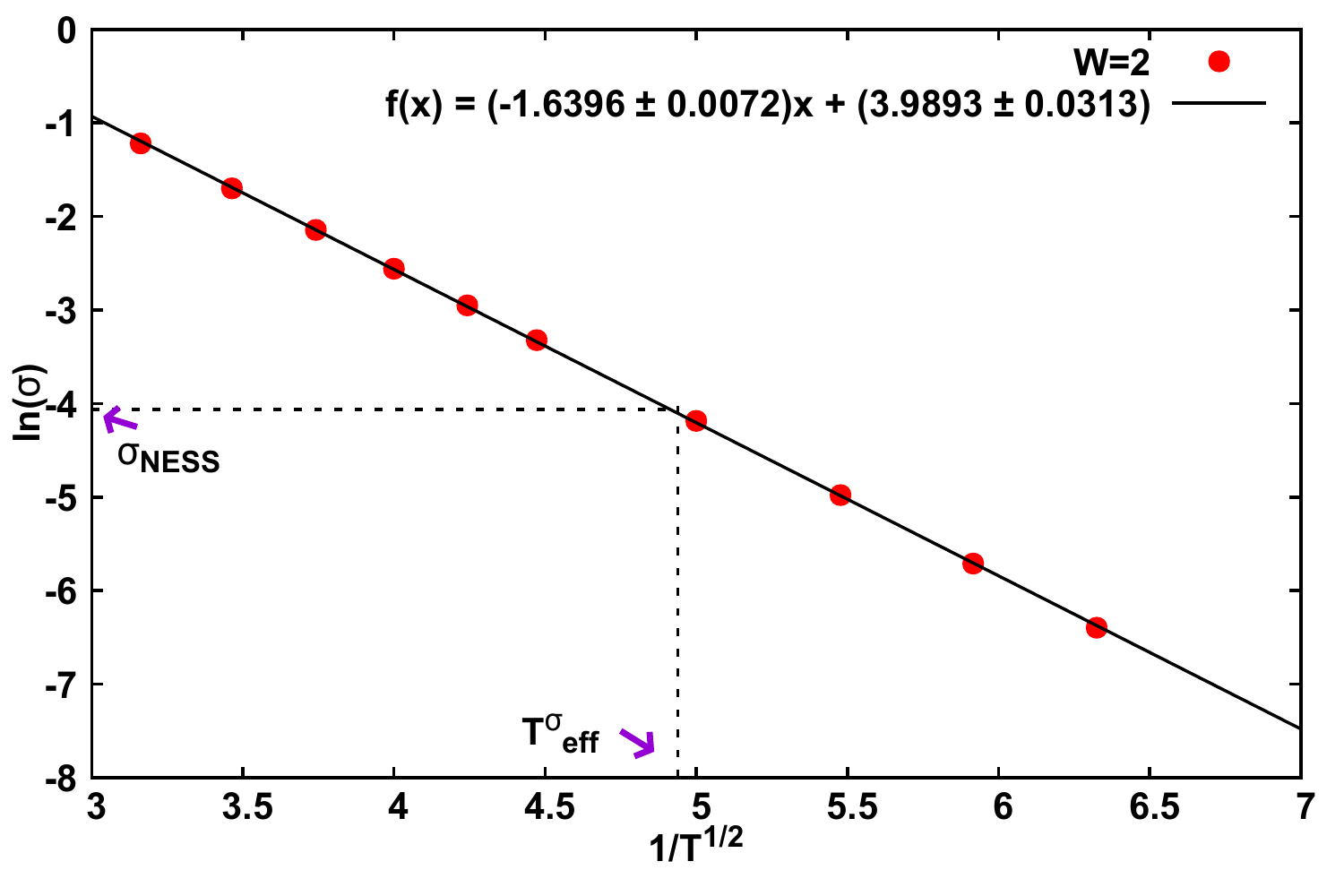}
    \caption{The conductivity of the thermal state as a function of temperature at $W = 2$. We calculate the effective temperature ($T_{eff}^{\sigma}$) of the NESS state here using the functional form $f(x)$ (the solid line giving the best fit here).}
    \label{fig:my_sigma}
\end{figure}

\underline{\textit{Using $g(0)$ vs $T$ relation}}: As shown in the main text (red circles in Fig.5)  and Fig.3 here, our data shows that the density of states at the Fermi-level $g(0)$ of the thermal states follows the following relation
\begin{equation}
    \label{dos_thermal}
    g(0) = c T^{\alpha}
\end{equation}
where $\alpha=1.29$ and $c$ is the proportionality constant. We calculate the effective temperature of the NESS state ($T_{eff}^{g}$) using the relation
\begin{equation}
    \label{dos_ness}
    g_{0}^{NESS}(x,T) = c \, (T_{eff}^{g})^{\alpha}
\end{equation}
where $g_{0}^{NESS}(x, T)$ is the density of states at the Fermi-level of the NESS state as a function of relaxation time $(x)$ and bath temperature $T$. 

\begin{figure}[h]
    \centering
    \includegraphics[scale=1.0]{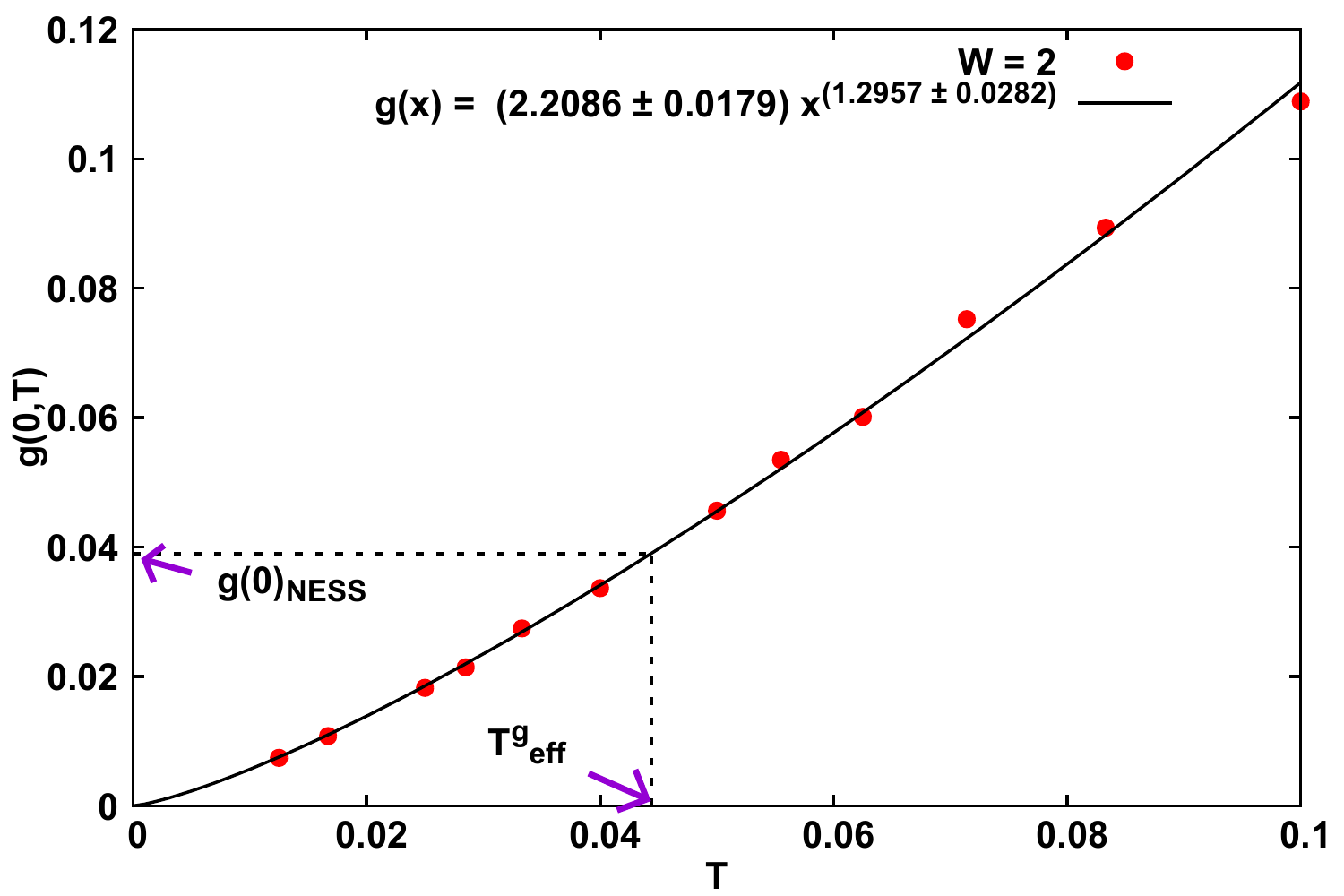}
    \caption{The density of states at the Fermi-level of the thermal state as a function of temperature at $W = 2$. We calculate the effective temperature ($T_{eff}^{g}$) of the NESS state here using the functional form $f(x)$ (the solid line giving the best fit here).}
    \label{fig:my_dos}
\end{figure}

\clearpage
\newpage
\underline{\textit{Comparison of the effective temperatures}}: Here, we compare the effective temperatures obtained using Eq.(\ref{FD_Teff}, \ref{sigma_ness} and \ref{dos_ness}) for various physical qualities at a constant temperature $T$ as a function of relaxation times ($x$).
\begin{figure*}[h]
\centering
\includegraphics[scale=0.55]{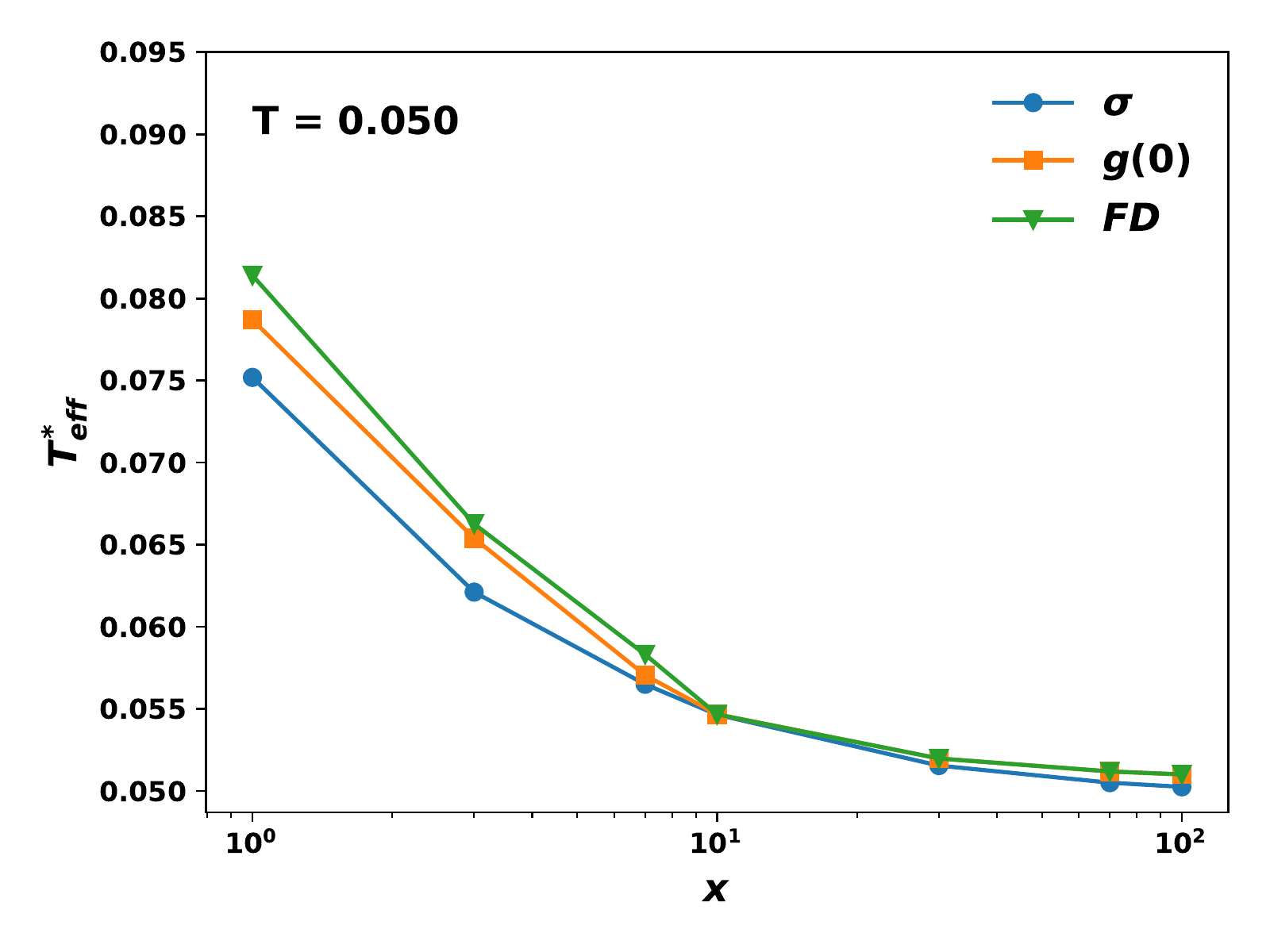}
\includegraphics[scale=0.55]{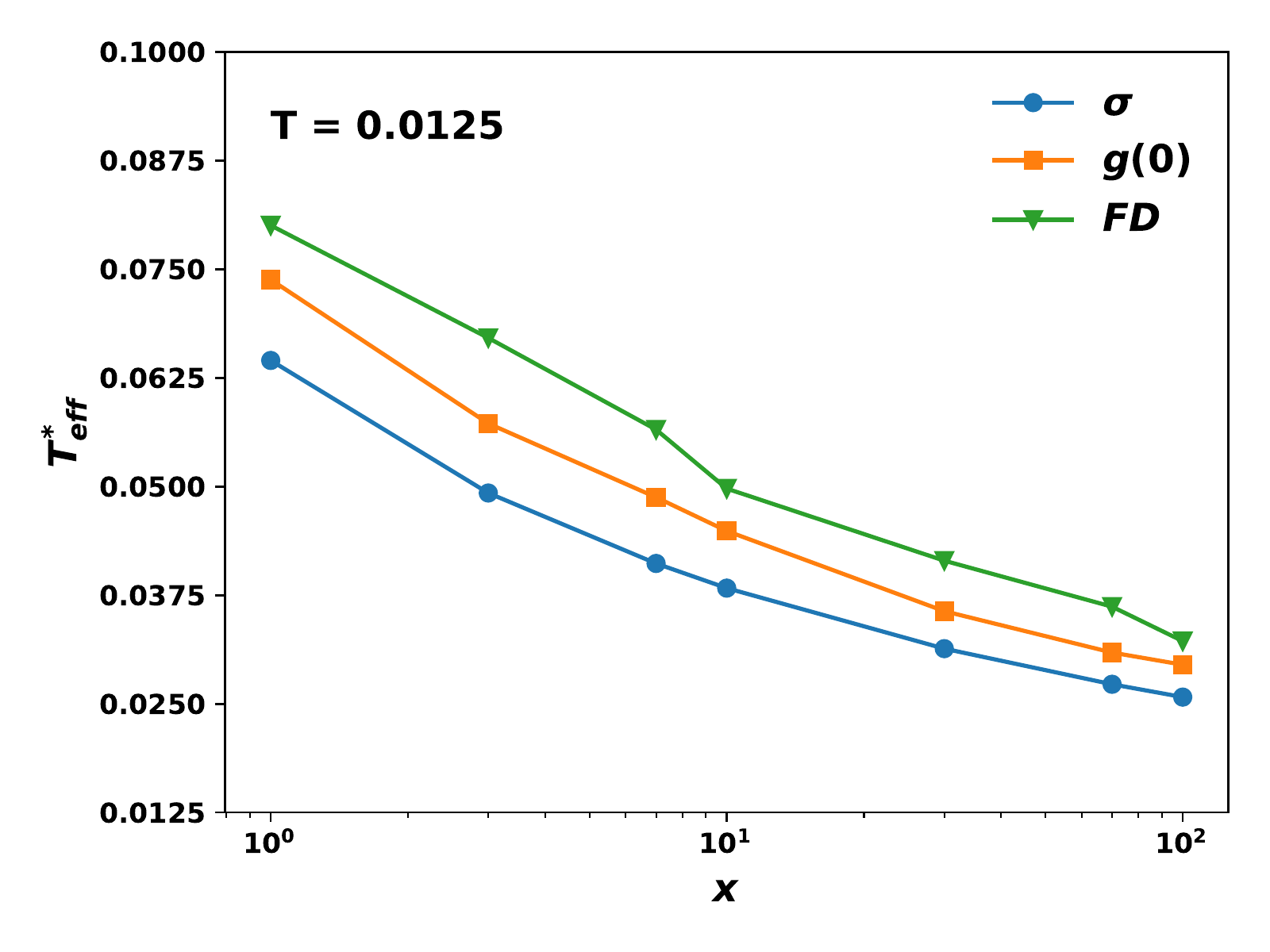}
\caption{\label{NESS_b80_Teff} The Fermi-Dirac distribution around the Fermi-level (FD), DOS at the Fermi-level ($g(0)$), and conductivity ($\sigma$) at the bath temperature $T$ were used to compute the effective temperature ($T_{eff}^{*}$) of the NESS state as a function of relaxation time ($x$). $T_{eff}^{*}$ here corresponds to $T_{eff}^{\sigma}$ for the blue line, $T_{eff}^{g}$ for the orange line and $T_{eff}^{FD}$ for the green line. }
\end{figure*}
\clearpage
\newpage

\section{Simulation results}

Supplementary table (\ref{table1}) presents the number of upward and downward transitions performed in the Monte Carlo simulations for this work.

For an equilibrium state, the number of downward transitions ($M^{\beta}_{d}$) is approximately equal to the number of upward transitions ($M^{\beta}_{u}$) (we have checked this in our simulations). We now denote $N^{\beta}_{u}(x)$ and $N^{\beta}_{d}(x)$ as the additional number of upward and downward transitions performed in the NESS state, respectively. Note that the total number of upward/downward transitions in a NESS state (n) is

\begin{equation}
 \label{MC_prob}
   n=\begin{cases}
    M^{\beta}_{d} + N^{\beta}_{d}(x) , & \Delta E \leqslant 0.\\
    M^{\beta}_{u} + N^{\beta}_{u}(x) , & \Delta E > 0.
  \end{cases} 
 \end{equation}
 
Two important points can be extracted from Table. (\ref{table1}): 

First, comparing Columns (3-4), we find that the number of downward transitions dominates over the upward transitions in the NESS state ($N^{\beta=80}_{d}(x=1) \textgreater\textgreater N^{\beta=80}_{u}(x=1) $). 
 
Second, if one compares $N^{\beta=80}_{d}(x=1)$ and $N^{\beta=20}_{d}(x=1)$, which corresponds to number of additional downward transitions at $\beta=80$ and $\beta=20$ respectively for relaxation time $x = 1$, we find that the two are approximately the same. This is true for $N^{\beta=40}_{d}(x=1)$ case as well (data not shown). This implies that $N^{\beta}_{d}(x)$ is temperature independent. Since these additional transitions are responsible for the change in conductivity ($\Delta \sigma$) of the system, $\Delta \sigma$ is temperature independent (this is shown in the inset of Fig.3 of the main text). 

\begin{table}[h]
\caption{\label{table1} Simulation parameters at temperature $\beta = 1/T$ at distance $r$ and relaxation time $x$ are presented. The data represents the number of Monte Carlo steps ($\times 10^{6}$) in downward and upward directions.}
\centering
\begin{tabular}{|l|l|l|l|l|}
 \hline
 distance & $M^{\beta=80}_{d}$ & $N^{\beta=80}_{u}(x=1)$ & $N^{\beta=80}_{d}(x=1)$ & $N^{\beta=20}_{d}(x=1)$   \\ \hline
\hline
$r = 1$  & 0.6460 & 0.8377 & 2.3203 & 2.0681  \\
\hline
$1 < r \leq 2$ & 0.1001 & 0.2248 & 1.4432 & 1.3959   \\
\hline
$2 < r \leq 3$ & 0.0194 & 0.0858 & 0.9415 & 0.9246  \\
\hline
$3 < r \leq 4$ & 0.0033 & 0.0290 & 0.3993 & 0.3932  \\
\hline
$4 < r \leq 5$ & 0.0008 & 0.0138 & 0.2076 & 0.2060  \\
\hline
$5 < r \leq 6$ & 0.0001 & 0.0047 & 0.0727 & 0.0719 \\
\hline
$6 < r \leq 7$ & 0.00004 & 0.0019 & 0.0307 & 0.0305  \\
\hline
$7 < r$        & 0.00002 & 0.0015 & 0.0246 & 0.0247  \\
\hline

\end{tabular}
\end{table}